\documentclass[runningheads]{llncs}
\usepackage[T1]{fontenc}
\usepackage[utf8]{inputenc}
\usepackage{newunicodechar}
\usepackage[USenglish]{babel}
\usepackage{amsmath}
\usepackage{amsfonts}
\usepackage{mathtools}
\usepackage{enumerate}
\usepackage{graphicx}
\usepackage{dashbox}
\usepackage{hyperref}

\DeclarePairedDelimiter{\floor}{\lfloor}{\rfloor}
\DeclarePairedDelimiter{\ceil}{\lceil}{\rceil}

\newunicodechar{≤}{\le}
\newunicodechar{≥}{\ge}
\newunicodechar{⌊}{\lfloor}
\newunicodechar{⌋}{\rfloor}
\newunicodechar{⌈}{\lceil}
\newunicodechar{⌉}{\rceil}
\newunicodechar{≈}{\approx}
\newunicodechar{·}{\cdot}
\newunicodechar{…}{\ldots}
\newunicodechar{½}{\frac{1}{2}}
\newunicodechar{−}{-}           

\newcommand{\Z}{\mathbb{Z}}

\newlength{\hfigsep}
\newlength{\vfigsep}
\newcommand{\s}[1]{\hspace{#1\hfigsep}}
\newcommand{\vs}[1]{\par\vspace{#1\vfigsep}\noindent\par}

\newcommand{\journal}[2]{#2}

\begin{document}

\title{Run-Length Encoding in a Finite Universe}
\author{N.\,Jesper Larsson\inst{1}}
\authorrunning{N.J. Larsson}
\institute{Malmö University, Faculty of Technology and Society,
  Malmö, Sweden}
\maketitle
\begin{abstract}
  Text compression schemes and compact data structures usually combine sophisticated probability models with basic coding methods whose average codeword length closely match the entropy of known distributions.
  In the frequent case where basic coding represents run-lengths of outcomes that have probability $p$, i.e.\@ the geometric distribution $\Pr(i)=p^i(1-p)$, a \emph{Golomb code} is an optimal instantaneous code, which has the additional advantage that codewords can be computed using only an integer parameter calculated from $p$, without need for a large or sophisticated data structure.
  Golomb coding does not, however, gracefully handle the case where run-lengths are bounded by a known integer~$n$. In this case, codewords allocated for the case $i>n$ are wasted.
  While negligible for large $n$, this makes Golomb coding unattractive in situations where $n$ is recurrently small, e.g., when representing many short lists of integers drawn from limited ranges, or when the range of $n$ is narrowed down by a recursive algorithm.

  We address the problem of choosing a code for this case, considering efficiency from both information-theoretic and computational perspectives, and arrive at a simple code that allows computing a codeword using only $O(1)$ simple computer operations and $O(1)$ machine words.
  We demonstrate experimentally that the resulting representation length is very close (equal in a majority of tested cases) to the optimal Huffman code, to the extent that the expected difference is practically negligible.
  We describe efficient branch-free implementation of encoding and decoding.
\end{abstract}

\section{Introduction}\label{sec:introduction}

We are concerned with efficiently computing short codewords for values that follow a geometric distribution up to a bounding integer. Before discussing application areas and outlining the nature of our result, we begin with a scenario that illustrates the precise nature of our problem.

\paragraph{A Context for the Problem}

In the original publication of a coding method for a geometric distribution, Golomb~\cite{Golomb} presents an introductory example involving Agent 00111 at the Casino, “playing a game of chance, while the fate of mankind
hangs in the balance.”
During 00111's game, the Secret Service wishes to continually receive status updates, and
has enrolled the bartender to communicate the length of every run of consecutive \emph{wins}, where a win is the more probable of two possible outcomes. Golomb presents his code as the solution to the problem of using the minimum number of bits in the transmission of run lengths.

To continue Golomb's scenario, we place ourselves in the situation of a competing agency that wishes to transmit the same information more efficiently, using the observation that there is always a finite upper bound~$n$ on the run length of consecutive wins, due to the need for 00111 to leave the game at predictable intervals to renew his cocktail. Since this means that only a finite number of possible run lengths exists, a \emph{Huffman code} for the different possible values would seem to be an option. But since $n$ varies, a new code would have to be computed for every codeword transmitted, and unlike Golomb coding, Huffman coding requires construction of a size $n$ data structure. To circumvent the Casino's scanning visitors for cheating devices, codewords must be computed by a computing unit woven into our agent's dress, with only a few registers of storage and minimal power consumption. Thus, in algorithmic terms, our mission, should we decide to accept it, is to find a method to compute short codewords for this scenario in $O(1)$ simple operations using $O(1)$ storage.

\paragraph{Application Areas}

Because of the prevalence of the geometric distribution in data, Golomb coding, and the computationally simplified version \emph{Rice coding}~\cite{Rice} are --~along with Huffman coding~\cite{Huf}~-- among the most common methods for low-level entropy coding in data compression.
Although arithmetic coding~\cite{Rissanen1976} can often produce better compression, it is inconvenient in many applications, e.g.\@ due to the necessity to decode previous parts of the bit sequence in order to access an arbitrary encoded value, which makes it less attractive for use in compact data structures~\cite{NavarroCompact}.

A common context for Golomb coding is representing lengths of gaps between events that are estimated to occur with some known small probability $1-p$. One common application is compressed representations of \emph{inverted index} data structures~\cite{MG,ZobelMoffatInvertedFiles2006}, but many others exist.
The significance of Golomb coding is demonstrated by a steady flow of new works citing Golomb~\cite{Golomb} as well as the variant devised and proven optimal by Gallager and van Voorhis~\cite{GallagerVoorhis}, which has become the standard formulation.\footnote{Golomb's formulation partitions values into groups of equal codeword length, which is a slightly less convenient take on the same family of codes.}
To take a small selection of examples, recent publications have included work on Golomb parameter estimation using machine learning~\cite{GParamEst}, use for biomedical data~\cite{GBiomedical}, lossless video compression~\cite{GLosslesVideo}, encoding phasor measurement to monitor the power grid~\cite{GPhasorMeas}, and replacements for Golomb and Rice coding for random access~\cite{KulekciRandomAccess}.


In the situation addressed in this work --~values to be encoded bounded by a relatively small integer~$n$~-- Golomb coding is unattractive (and may therefore fail to be considered a possibility), because the part that the Golomb code dedicates to values above $n$ would be wasted, rendering redundancy unnecessarily large.

In the compressed-inverted-index context, this could arise when document references in many (presumably small) lists are drawn from a limited collection.
More typically, the upper bound on run length appears because integers are encoded recursively, contained in intervals that shrink with recursion depth, similarly to schemes such as interpolative coding~\cite{Interpolative}, tournament coding~\cite{TeuholaTournament}, wavelet trees~\cite{GrossiGuptaVitter}, or recursive subset coding~\cite{LarssonSetcompArxiv}.

An available efficient entropy coding method for the bounded geometric distribution can contribute to development of effective compression schemes that do not yet exist.

\journal{9) Page 12, Section 6: "can plausibly contribute to the design of efficient text compression algorithms and succinct data structures". I have a couple of suggestions about papers that use Golomb-compressed data structures:

  - The following paper from 2007 deals with Golomb codes and succinct data structures:

  O'Neil Delpratt, Naila Rahman, Rajeev Raman: Compressed Prefix Sums. SOFSEM (1) 2007: 235-247.

  It would be good to comment about this, and relate it to the proposed result.

  - The following paper:

  Hao Yan, Shuai Ding, Torsten Suel: Compressing term positions in web indexes. SIGIR 2009: 147-154

  defines adaptive variants of Golomb/Rice codes. These are variants of Rice codes able to encode small values, corresponding to position information in inverted indexes. Would the proposed result be good for this application?}{}

\paragraph{Outline}

We propose a method that computes a codeword for the run length of outcomes that have a known probability~$p≥½$,
given an upper bound~$n$, and two integer parameters~$m$ and $m''$ chosen on the basis of $p$. The method uses only a constant number of machine words, and a constant number of simple computational operations per computed codeword. The expected codeword length is close to minimum.


We generally assume (although our coding does not strictly depend on it) that only a small number of different $p$ --~and hence a small collection of $m$ and $m''$~-- appear in processing a single file or data structure. Therefore, if $m$ and $m''$ need to be included in the compressed representation (because $p$ is not available to the decoder), we assume that the encoding length as well as encoding time is negligible. Furthermore, we assume that $n$ is available from the context to both encoder and decoder, and does not need to be explicitly represented.
See section~\ref{sec:efficiency} for a more detailed account of time complexity, along with a description of an efficient branch-free implementation. 

Section~\ref{sec:preliminaries} sets the scene by relating theory and relevant previous methods, and section~\ref{sec:thecode} presents our suggested code. Section~\ref{sec:eval} evaluates compression performance, section~\ref{sec:efficiency} adresses time complexity and efficient implementation, and we conclude with some short comments in section~\ref{sec:conclusion}.

\section{Entropy Codes and Code Trees}\label{sec:preliminaries}

Let $X$ be a random variable that takes on non-negative integer values $i\in{0,…,n}$ with probabilities $\Pr(X=i)=p_i$.
We are concerned with finding codes that map the possible values of $X$ to binary codewords of minimum expected length.
A lower bound is the \emph{entropy} $H(X)=-\sum_{i=0}^n p_i\lg p_i$ bits, where $\lg$ denotes logarithm in base~2.
The bound can be matched only if all probabilities are \emph{dyadic}, i.e., $\lg p_i\in\Z$. The \emph{redundancy} of a code is the difference between its expected codeword length and the entropy.
Any \emph{instantaneous} code, where no codeword is a prefix of another, allows us to equate each code with a \emph{code  tree}: a binary tree with all possible values as leaves, and codewords identified by their paths from the root ($0$ for left branching, $1$ for right branching).
We refer to the number of values in a code tree (i.e.\@ the number of leaves) as the \emph{size} of the tree.
It is well known that a \emph{Huffman tree}, constructed bottom up by repeatedly joining the two subtrees of lowest probability, is optimal, i.e., no code that maps an individual value to each codeword can have shorter expected codeword length.~\cite{CoverThomas,MG}

\begin{figure}
  \begin{minipage}{1\textwidth}
    \footnotesize
    \setlength{\parsep}{0pt}
    \setlength{\parskip}{0pt}
    \setlength{\hfigsep}{1.33em}
    \setlength{\vfigsep}{3.0ex}
    \hfil\includegraphics[scale=.8]{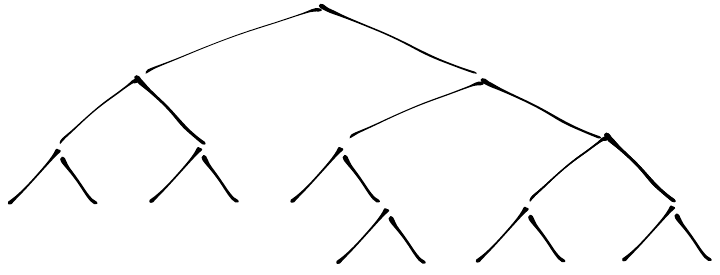}\hfil\raisebox{-3ex}{\includegraphics[scale=.8]{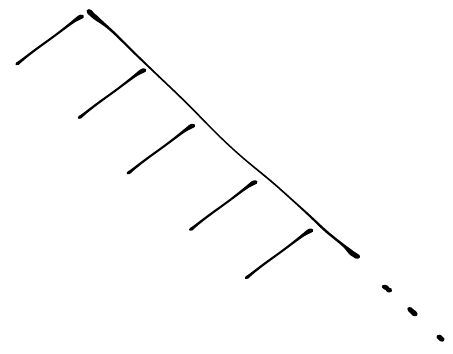}}\hfil
    \vs{-2.1}
    \s{2.1}$0$\s{1.2}$1$\s{.7}$2$\s{1.2}$3$\s{.6}$4$
    \vs{.2}
    \s{2.1}\s{6.2}$5$\s{1.2}$6$\s{.7}$7$\s{1.2}$8$\s{.7}$9$\s{1.1}$10$
    \vs{-5.6}
    \s{17.7}$0$\vs{.15}
    \s{17.7}\s{1.2}$1$\vs{.15}
    \s{17.7}\s{2.1}$2$\vs{.15}
    \s{17.7}\s{3.2}$3$\vs{.15}
    \s{17.7}\s{4.3}$4$\vs{1}
  \end{minipage}
  \caption{Balanced code tree for $n=10$ (left), and unlimited unary code tree (right).} \label{fig:balanced}\label{fig:unary}
\end{figure}

When probabilities are roughly equal, or more specifically, the highest probability is no greater than the sum of the two lowest probabilities, codeword lengths in an optimal code differ by at most one bit.
Therefore, an optimal code is achieved by using codeword length $h-1$ for the $s$ lowest-probability values and $h$ for the other $n+1-s$, where $h=⌈\lg (n+1)⌉$ and $s=2^h-(n+1)$ (see figure~\ref{fig:balanced}).
We refer to this as a \emph{balanced code}.
\footnote{Many authors use the term \emph{binary code}. However, this clashes with the fact that all codes we address are binary in a wider sense.}

On the other extreme, if the~$p_i$ decrease at least as quickly as proportionally to a reverse Fibonacci sequence, an optimal code is achieved by a code tree that is as skewed as possible, e.g.\@ by encoding $i$ as $i$ one-bits followed by a single zero~bit.
This is referred to as a \emph{unary} code (figure~\ref{fig:unary}). With an upper bound $n$ for $i$, the final zero in the codeword for $n$ can be excluded.

When $p_i=p^i(1-p)$ (the geometric distribution) for some $½≤ p < 1$, $i≥0$, and $p^m=½$ for some integer $m$, an optimal code is achieved by partitioning values into groups of size $m$, which we shall refer to as \emph{bunches} in the code tree, and composing the code from two parts: a unary code that numbers the bunch of each value counting from the root, and a balanced code that determines the offset of each value inside its bunch.
This is referred to as a \emph{Golomb code} (figure~\ref{fig:golomb}).
More generally, let $m$ be the integer such that $p^m$ is as close as possible to $½$.
If $p^m\ge½$, then $p^{m+1}<½$, and $m$ is the smallest integer such that $p^m-½≤½-p^{m+1} \Leftrightarrow p^m+p^{m+1}≤ 1$.
If $p^m≤½$, $m$ is again the smallest integer for which $p^m+p^{m+1}≤1$.
Consequently,
\begin{equation}\label{eq:bunch}
  m = \min\left\{\,\ell\in\Z \;\bigm|\; p^\ell + p^{\ell+1}≤ 1\,\right\} = \ceil*{\frac{\lg(1+p)}{-\lg p}}
\end{equation}
Note that if $p=½$, then $m=1$, and the Golomb code is equal to a unary code.

Gallager and van Voorhis~\cite{GallagerVoorhis} showed, using a bottom-up Huffman tree argument, that choosing $m$ according to equation~\ref{eq:bunch} always produces an optimal code tree. An alternative extended treatment was given by Golin~\cite{Golin}.

\section{Finite code trees for the geometric distribution}\label{sec:thecode}

We now turn to the problem of forming code trees for a geometric distributions up to a maximum value~$n$:

\begin{align*}
  \Pr(X=i) &= p^i(1-p) \text{~for~} i\in\Z \text{~and~} 0≤ i< n\text{,} \\
  \Pr(X=n) &= 1-\sum_{i=0}^{n-1}p^i(1-p) = p^n \text{, and} \\
  \Pr(X=x )&=0 \text{~for all other~} x.
\end{align*}
We write $H(p, n)=-\sum_{i=0}^n p^i(1-p)\lg (p^i(1-p))$ for the entropy under this distribution, and denote the expected code length for a corresponding optimal code tree (e.g.\@ Huffman) $L_H(p, n)\ge H(p, n)$


For convenience in encoding and decoding, we require values $0,…,n$ to be in left to right order in the code tree, which does not impose any penalty on codeword lengths.
To see this, consider starting from the \emph{canonical} code tree~\cite{SchwarzKallick,MG}, where codeword lengths are optimal, and leaves have monotonically non-decreasing depth from left to right.
This differs from our desired code tree only in the placement of leaf $n$, which does not generally have the smallest codeword length, and is therefore not generally the rightmost leaf in the canonical tree.
For every internal node in the path from the root to $n$, exchange the left and right subtrees wherever $n$ is in the left subtree.
This places $n$ as the rightmost leaf, while maintaining the depths of all leaves.
Hence, exchanging subtrees can displace the order only among leaves with the same depth, and since all leaves other than $n$ have strictly decreasing probability order, they can be set in the right order without compromising the optimality of the tree.


\subsection{Approach}\label{sec:approach}

When $n\rightarrow\infty$, a Golomb code tree is optimal~\cite{GallagerVoorhis}, and a fair guess is that the code for finite $n$ has some similarity to a Golomb code.
Indeed, since the wasted probability $p^n$ is negligible for large $n$, a hybrid scheme that uses Huffman coding if $n$ is smaller than some acceptable constant, and Golomb coding otherwise, would theoretically solve our problem. However, in the interest of practical efficiency and simplicity, we seek a coding method that operates uniformly for all~$n$.

\begin{figure}
  \footnotesize
  \setlength{\parsep}{0pt}
  \setlength{\parskip}{0pt}
  \setlength{\hfigsep}{1.33em}
  \setlength{\vfigsep}{3.0ex}
  \begin{minipage}[t]{0.49\textwidth}
    \null
    
    \begin{center}
      \includegraphics[scale=.8]{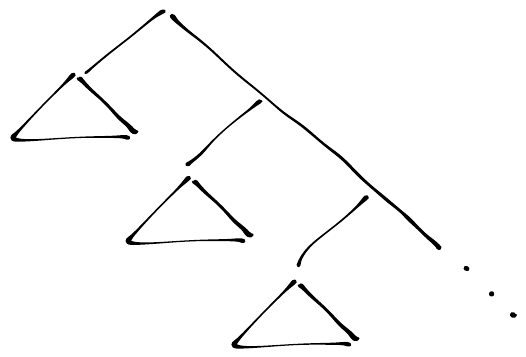}
    \end{center}
    \vs{-6.4}\s{3.2}$m$
    \vs{1.1}\s{5.35}$m$
    \vs{1.4}\s{7.3}$m$
    {\scriptsize\vs{-4}\s{2.2}$0$\s{1.5}$m\!\!-\!\!1$}
    {\scriptsize\vs{1.3}\s{4.2}$m$\s{1.2}$2m\!\!-\!\!1$}
    {\scriptsize\vs{1.1}\s{6.2}$2m$\s{1}$3m\!\!-\!\!1$}

    \null
  \end{minipage}
  \begin{minipage}[t]{0.49\textwidth}
    \null
    
    \begin{center}
      \includegraphics[scale=.8]{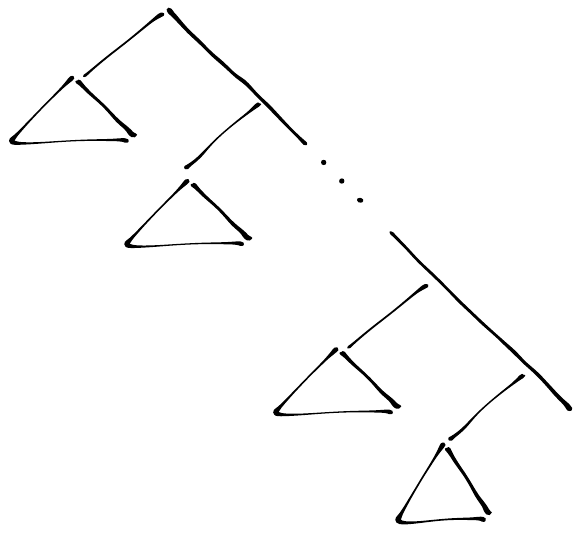}
    \end{center}
    \vs{-9.9}\s{2.7}$m$
    \vs{1.1}\s{4.8}$m$
    \vs{2.5}\s{7.5}$m$
    \vs{1.1}\s{9.5}$m_1$

    {\scriptsize\vs{-7.3}\s{1.7}$0$\s{1.5}$m\!\!-\!\!1$}
    {\scriptsize\vs{1.3}\s{3.8}$m$\s{1.2}$2m\!\!-\!\!1$}
    {\scriptsize\vs{4.5}\s{8.2}$n\!\!-\!\!m_1$\s{.6}$n\!\!-\!\!1$}
    \vs{-3.2}\s{12.4}$n$
    \vs{2}
    
    \null
  \end{minipage}
  \caption{Golomb code tree (left), and top-down weight balanced tree (right). Triangles are balanced code subtrees with sizes (number of leaves) shown inside and value ranges below.} \label{fig:golomb}\label{fig:naivewb}
\end{figure}

A first attempt might be to use a Golomb code with minimal modification: Let $m$ be defined by equation~\ref{eq:bunch}, and define $d_1=\floor*{\frac{n-1}{m}}$ and $m_1=n-d_1m≤m$.
Assign the first $d_1m$ values the same codewords as in the corresponding Golomb code.
Let the rightmost internal node at depth~$d_1$ have leaf~$n$ as its right child,  and a balanced code tree for values $n-m_1,…,n-1$,  as its left subtree (figure~\ref{fig:naivewb}).
Consider the nodes along the rightmost path of the resulting tree.
In general, the internal node at depth $k$ along this path ($0≤ k< d_1$) separates values into two subtrees with probability weights $\sum_{i=km}^{(k+1)m-1}p^i(1-p)=p^{km}-p^{(k+1)m}$ and $p^{(k+1)m}$.
Dividing by $p^{km}$, we see that the relative weights of left and right subtree are $1-p^m$ and $p^m$, which given our definition of $m$ implies (unsurprisingly) that the subtrees are as weight-balanced as possible.
The final internal node on the rightmost path has a left subtree of size~$m_1$ (with probability weight $p^{n-m_1}-p^n$), and leaf $n$ for right child (with weight $p^n$).
This implies that at least locally, this node also balances the weights among its descendents as equally as possible.

Consequently, this tree is similar to what would be obtained by the top-down method known as Shannon--Fano coding, or more precisely (since our leaf probabilities are not necessarily monotonically non-increasing), a weight-balanced code tree as studied by Rissanen~\cite{RissanenWeightBalanced} and Horibe~\cite{Horibe}.
The expected codeword length is therefore guaranteed to be asymptotically close to $H(p, n)$, but not necessarily optimal.
In particular, note the striking anomaly when $m_1$ is much smaller than $m$, that a shorter codeword may be assigned to $n-m_1$ (and possibly others) than to $n-m_1-1$.
This causes substantial redundancy for some $p$ and~$n$.
For instance, when $p=0.88$ and $n=6$, the expected codeword length is $2.63$, ten percent more than the optimal $L_H(0.88, 6)=2.38$.\label{sec:optimaliscomplex}

The reason is that greedy top-down partitioning can leave us with a severe unbalance between leaf $n$ and the size-$m_1$ subtree, while a Huffman tree construction, which is bottom-up, can distribute weight among bunches of lower depth for better overall balance.
For example, with $p=0.9$ and $n=43$, the optimal tree has four bunches of size~$m=7$ and one of size~$8$. With the same $p$ and $n=44$, only two size $m$ bunches emerge, three of size $6$, and one of size~$5$.

It is unclear whether a rule can be found that assigns bunch sizes optimally while allowing any codeword to be computed without explicitly processing $\omega(1)$ nodes in the code tree.
Therefore, rather than computing the optimal code, we propose the following coding, which renders codewords almost as easy to compute as the top-down weight balanced code trees.

\subsection{Suggested Code}\label{sec:algorithms}

Our suggested code trees are shaped similarly to the corresponding Golomb trees, with the exception of a subtree at the bottom that we refer to as the \emph{tail}. Unless $n<m$, the tail has at least $m+1$ and at most $2m$ leaves, whose rightmost leaf is~$n$, and whose other leaves have as equal depths as possible (similarly to a balanced code). If $n<2m$, the whole tree is comprised of the tail.

Encoding and decoding takes two parameters~$m$ and $m''$, where $m$ should be chosen according to equation~\ref{eq:bunch}, and the choice of integer $m''$, $m<m''≤2m$, is discussed at the end of this section and defined in equation~\ref{eq:mbis}. Define
\begin{align*}
  m'&=\min\{m+(n\bmod m), n\}\text{, and} \\
  d_t&=\frac{n-m'}{m}\text{.}       
\end{align*}
Note that $\min\{n, m\}≤ m'<2m$, and that $d_t$, the \emph{tail depth}, is an integer.

The codeword for $i$ where $0≤ i < d_tm$ consists of $⌊i/m⌋$ one-bits followed by the binary representation for $i\bmod m$ left-padded with zeros up to $h=⌈\lg m⌉$  bits if $i\bmod m<2^h-m$, and up to $h+1$  bits otherwise. This is equivalent to a unary code for the bunch number followed by a balanced code for the offset inside the bunch, i.e., the same as in the corresponding Golomb code.

\begin{figure}{
    \setlength{\parsep}{0pt}
    \setlength{\parskip}{0pt}
    \setlength{\dashlength}{3.5pt}
    \setlength{\dashdash}{.75pt}
    \footnotesize
    \setlength{\hfigsep}{1.33em}
    \setlength{\vfigsep}{3.0ex}
    \begin{minipage}[c]{0.49\textwidth}
      \null
      
      \begin{center}
        \includegraphics[scale=.8]{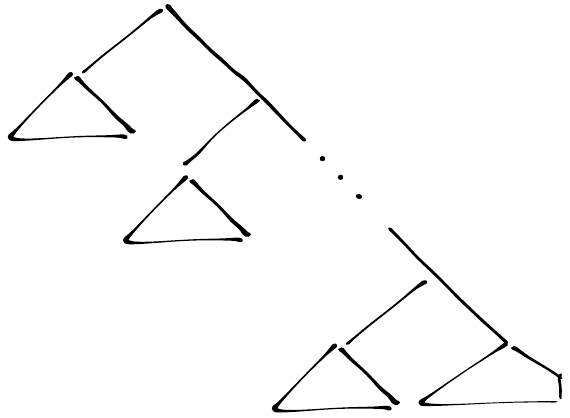}
      \end{center}
      \vs{-9.3}
      \s{4.8}
      \vbox{
        \hbox{\rule{7.4\hfigsep}{0.125pt}}
        \hbox{\s{7.3}$\left.\rule{0pt}{3.6\vfigsep}\right\}d_t$}
        \hbox{\s{6.4}\rule{1\hfigsep}{.125pt}}
      }
      \vs{-5.3}\s{2.7}$m$
      \vs{1.1}\s{4.85}$m$
      \vs{2.65}\s{7.5}$m$\s{2.0}$m'\!\!+\!\!1$
      {\scriptsize\vs{-5.2}\s{1.65}$0$\s{1.5}$m\!\!-\!\!1$}
      {\scriptsize\vs{1.3}\s{3.55}$m$\s{1.5}$2m\!\!-\!\!1$}
      {\scriptsize\vs{2.5}\s{6.35}$d_tm$\s{1.95}$n\!\!-\!\!m'$\s{1}$n$}
      \vs{-2.6}\s{9.3}\dashbox{\rule{3.1\hfigsep}{0pt}\rule{0pt}{2.1\vfigsep}}
      \vs{0}\s{12}{\scriptsize tail}
      
      \null
    \end{minipage}
    \begin{minipage}[c]{0.49\textwidth}
      \null
      
      \begin{center}
        \includegraphics[scale=.8]{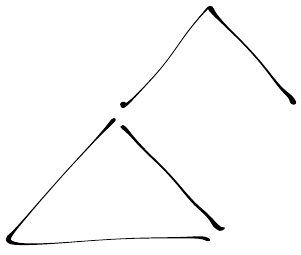}
        \vs{1.5}
        \includegraphics[scale=.8]{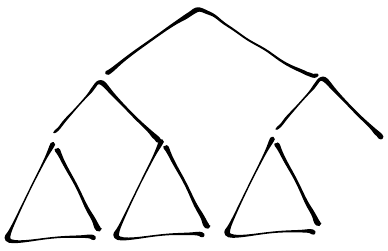}
      \end{center}
      \vs{-12.5} 
      \vbox{
        \hbox{\s{3.2}\rule{4.8\hfigsep}{0.125pt}}
        \hbox{\s{1.8}$h'\left\{\rule{0pt}{2.5\vfigsep}\right.$}
        \hbox{\s{3.2}\rule{1.1\hfigsep}{.125pt}}
      }
      \vs{1.7} 
      \vbox{
        \hbox{\s{2.2}\rule{4.8\hfigsep}{0.125pt}}
        \hbox{\s{0.8}$h'\left\{\rule{0pt}{2.5\vfigsep}\right.$}
        \hbox{\s{2.2}\rule{1.1\hfigsep}{.125pt}}
      }
      \vs{-8.3}\s{6}$m'$
      \vs{-2.4}\s{9.8}$n$
      \vs{1.8}\s{4}$n\!-\!m$\s{1.8}$n\!-\!1$
      \vs{3.6}\s{10.5}$n$
      \vs{.08}
      \s{3.75}{\scriptsize{\tiny$\sim$}$\!\frac{m'}{3}$}
      \s{.55}{\scriptsize{\tiny$\sim$}$\!\frac{m'}{3}$}
      \s{.6}{\scriptsize{\tiny$\sim$}$\!\frac{m'}{3}$}
      \vs{.1}\s{3}$n\!-\!m$\s{3.7}$n\!-\!1$
    \end{minipage}
  }
  \caption{General code tree for suggested coding. Left, the overall code tree. Top right, the tail for case $e_n=1$, and bottom right for case $e_n=2$.} \label{fig:fgcode}
\end{figure}

The encoding of the remaining $m'+1$ values (the tail) depends on three parameters: the height of the tail subtree, which we denote $h'$;
the depth of $n$ within the tail subtree, which is either $1$ or $2$ and denoted $e_n$;
and the number of shorter codewords among the values $n-m',…,n-1$, denoted $s'$ (cf.\@ $s$ for the $m$-sized bunches).
The values are assigned as follows:
\begin{itemize}
\item If $m' < m''$, we let $e_n=1$, $h'=⌈\lg m'⌉+1$, and $s'=2^{h'-1}-m'$ (depicted top right in figure~\ref{fig:fgcode}).
\item Otherwise, $e_n=2$, $h'=\ceil*{\lg \frac{4m'}{3}}$, and $s'=3·2^{h'-2}-m'$ (depicted bottom right in figure~\ref{fig:fgcode}). Our choice of $m''$ will ensure that this case arises only when $m'≥3$.
\end{itemize}
The codeword for $i$ when $n-m'≤ i < n$ is $d_t$ one-bits followed by the binary representation of $j=i-m·d_t$, left-padded with zeros up to $h'-1$ bits when $j<s'$, and up to $h'$ bits when $j\ge s'$. Finally, the codeword of $n$ is $d_t+e_n$ one-bits. A demonstration implementation of encoding and decoding in Python can be found in appendix~1.

\subsubsection{Correctness and Decoding}

Our encoding corresponds to the general code tree shown in figure~\ref{fig:fgcode}. Therefore, given $m$, $m''$ and $n$, it yields a well-defined instantaneous code for integers $i$ from $0$ to~$n$.

Decoding reads one-bits from the input, stopping after a zero-bit has appeared or $d_t$~bits have been read.
The number of one-bits found, which we denote~$d_i$, is the number of an $m$-sized bunch if $d_i<d_t$, in which case decoding procedes with reading the offset in the bunch as an ordinary balanced code.
Otherwise, we interpret the following $h'$ bits as a number~$j$ that identifies a leaf in the tail.
(If less than $h'$ bits remain in the input, we right-pad with any bits.)
If the $e_n$ most significant bits of $j$ are one, the decoded value is~$n$. Otherwise, if $j<2s$ we are in the range of the shorter codewords, which implies that $h'-1$ bits should be consumed and the decoded value of $i$ is $m·d_t+⌊\frac{j}{2}⌋$; and if $j≥2s$, $h'$ bits should be consumed and $i=m·d_t+j-s'$

\subsubsection{Justification and Choice of $m''$}

If $m$ divides $n$ and $n≥m$, then $m'=m$. Since we stipulated that $m''>m$, the shape of our code tree in this case is the same as the optimal tree for unbounded values~\cite{GallagerVoorhis} except that the subtree for values above $n-1$, with total probability $p^n$, is contracted into a single leaf.
Since the code tree of Gallager and van Voorhis is optimal, our tree must also be optimal for $m'=m$, since otherwise the subtree that our code contracts would have had to be placed differently in the optimal unbounded tree.\label{sec:optimalm}

Consequently, it is clear that the ideal value of $e_n$ (the local depth of $n$ in the tail subtree) is $1$ when $m'=m$.
On the other hand, we can see that as $m'$ approaches $2m$, the ideal value of $e_n$ should ultimately be $2$.
To decide where in the range between $m$ and $2m$ to place the limit $m''$ between cases $e_n=1$ and $e_n=2$, we compare the expected codeword lengths for the two tail variants in figure~\ref{fig:fgcode}. The codeword lengths for values below~$n$ are between $⌊\lg m'⌋ $ and $⌈\lg m'⌉$ in the $e_n=1$ case, and between $2+⌊\lg \frac{4m'}{3}⌋ $ and $2+⌈\lg \frac{4m'}{3}⌉$ in the $e_n=2$ case. Using the respective approximation $\lg m'$ for the codeword lengths in the $e_n=1$ case and $2+\lg\frac{4m'}{3}$ in the $e_n=2$ case, we estimate that $e_n=1$ results in a smaller expected codeword length than $e_n=2$ when
\begin{equation*}
  (p^{n-m'}-p^n)(1+\lg m')+1·p^{m'} < p^{n}(2+\lg \frac{m'}{3})+2·p^{m'}\text{.}
\end{equation*}
Simplifying and solving for $m'$ yields
\begin{equation*}
  m' < \frac{\lg(\frac{1}{\lg 3 - 1} + 1)}{-\lg p} ≈ \frac{1.4380}{-\lg p}\text{.}
\end{equation*}
Hence, it is reasonable to choose $m''$, the value that determines the value of $e_n$ in our code specification, as
\begin{equation}\label{eq:mbis}
  m'' = \ceil*{\frac{1.4380}{-\lg p}}\text{.}
\end{equation}

We note that this implies that $m''=2$ only when $m=1 \Rightarrow m'=1$, which avoids the degenerate case where $e_n=2$ coincides with $m'<3$, necessary for our code tree to be well-defined.

The choice of $m''$ is a heuristic that is not guaranteed to result in an optimal code tree. The same is true for our choice to assign codewords to $n-m',…,n-1$ whose lengths differ by at most one: a difference of two (but never more than two, by the choice of~$m'$) may result in a slightly superior code. Also, as noted in section~\ref{sec:optimaliscomplex}, our choice of placing all but the last $m'$ values in size-$m$ bunches is not always optimal. Rather, these choices were made in order to obtain easily computable codewords without deviating much from the optimum. The next section evaluates to what extent this is successful.

\section{Evaluation of Code Efficiency}\label{sec:eval}

Let $L(p, n)$ be the expected codeword length of our code for specified $p$ and $n$. Using a bound derived by Horibe~\cite{Horibe}, we can conclude that $L(p, n)≤H(p, n)+2-(n+1)p^{n-1}(1-p)$. However, as worst-case bounds on codeword length tend to be, this is pessimistic, and not a useful realistic estimate of code performance.

For a more practically valid assessment, we first note that our codes are proven optimal when $m=m'$, i.e., for every $m$th value of $n$ (as we saw in section~\ref{sec:optimalm}).
For other values of $n$, our codes may deviate slightly from the optimal, but we have made heuristic choices so as to stay as close as possible, without sacrificing the possibility of efficiently computing codewords.
In particular, we note that our choice to assign values $n-m',…,n-1$ codewords whose lengths are monotonically nondecreasing and differing by at most one bit, eliminates the anomaly of the top-down weight-balanced tree that a less probable value may be assigned a shorter codeword (see section~\ref{sec:approach}).
Furthermore, we note that both our code and an optimal Huffman code approach a Golomb code for relatively large~$n$.
For instance, when $p=0.9$, the redundancy of our code, a Huffman code, and a Golomb code are all below one percent when $n≥54$.

Therefore, our main concern is to evaluate the redundancy of our code for smaller~$n$. We experimentally compare our code to the calculated entropy (which can be matched in practice only by high-precision arithmetic coding) as well as to the optimal Huffman code, when $p$ varies across its probability range and $n<3m$.
This range of~$n$ is the most interesting, since it results in at most one $m$-sized bunch plus the tail.
For reference, we also include a comparison with Golomb coding for the same test cases.

Tables~\ref{tab:entropy}–\ref{tab:golomb} summarize the results of tests for~$10^7$ values of $p$ evenly spread over the interval $½≤p<1$.
For each value of $p$, 10 random values of $n$ were selected uniformly at random in the range $[2,3m)$ (the case $n=1$ is uninteresting) for the comparison with Huffman (table~\ref{tab:huffman}) and Golomb codes (table~\ref{tab:golomb}). For the comparison with the entropy (table~\ref{tab:entropy}), the range $\left[\max\{2, ⌈\frac{m}{2}⌉\}, 3m\right)$ was used, because the entropy for smaller $n$ tend to be impossible to achieve with any code tree, and therefore less interesting for our evaluation. Python source code used for measurements can be found at \url{http://fgcode.avadeaux.net/}.

\begin{table}
  \caption{Evaluation in relation to entropy. The first column is the high endpoint of the redundancy range, e.g., we have 1–2 percent redundancy in 33.0 percent of the test cases. The $p$ and $n$ columns are examples that produce results in the respective range.}\label{tab:entropy}
  \begin{center}
    \renewcommand{\arraystretch}{1.05}
    \begin{tabular}{lr@{\hskip 1em}r@{,\,}l}
      high $(L-H)/H$ & \% & sample\hspace{.5em}$p$& $n$ \\
      \hline
      $0$ & 0.0 & $0.5$ & $2$\\
      $10^{-5}$ & 0.4 & $0.501$ & $2$\\
      $10^{-4}$ & 0.8 & $0.502$ & $2$\\
      $0.001$ & 2.5 & $0.506$ & $2$\\
      $0.005$ & 7.1 & $0.974$ & $52$\\
      $0.01$ & 26.7 & $0.984$ & $81$\\
      $0.02$ & 33.0 & $0.987$ & $87$\\
      $0.03$ & 13.0 & $0.919$ & $12$\\
      $0.05$ & 14.1 & $0.988$ & $34$\\
      $0.1$ & 1.9 & $0.983$ & $20$\\
      $0.5$ & 0.6 & $0.994$ & $45$\\
      $>0.5$ & 0 & \multicolumn{2}{c}{}\\
    \end{tabular}
  \end{center}
\end{table}

\begin{table}
  \caption{Evaluation in relation to Huffman coding. The first column is the high endpoint of the length increase range, e.g., our code yields output 1–2~percent larger than optimal in 0.3 percent of our test cases, and optimal in 86.2 percent of the cases. The $p$ and $n$ columns are examples that produce results in the respective range.}\label{tab:huffman}
  \begin{center}
    \renewcommand{\arraystretch}{1.05}
    \begin{tabular}{lr@{\hskip 1em}r@{,\,}l}
      high $(L-L_H)/L_H$ & \% & sample\hspace{.5em}$p$& $n$ \\
      \hline
      $0$ & 86.2 & $0.985$ & $93$\\
      $10^{-5}$ & 0.1 & $0.979$ & $68$\\
      $10^{-4}$ & 0.6 & $0.992$ & $175$\\
      $0.001$ & 4.1 & $0.972$ & $62$\\
      $0.005$ & 7.1 & $0.971$ & $67$\\
      $0.01$ & 1.6 & $0.938$ & $21$\\
      $0.02$ & 0.3 & $0.904$ & $12$\\
      $>0.02$ & 0 & \multicolumn{2}{c}{}\\
    \end{tabular}
  \end{center}
\end{table}

\begin{table}
  \caption{Comparison to Golomb coding. The first column is the high endpoint of the length decrease range, e.g., our code yields output at least 5 percent shorter than Golomb codes for all test cases, and 10–50 percent reduction in 84.2 percent of the cases. The $p$ and $n$ columns are examples that produce results in the respective range.}\label{tab:golomb}
  \begin{center}
    \renewcommand{\arraystretch}{1.05}
    \begin{tabular}{lr@{\hskip 1em}r@{,\,}l}
      high $(L_{\text{Golomb}}-L)/L_{\text{Golomb}}$ & \% & sample\hspace{.5em}$p$& $n$ \\
      \hline
      $0.05$ & 0 &  \multicolumn{1}{c}{} \\ 
      $0.1$ & 7.5 & $0.862$ & $14$\\
      $0.5$ & 84.2 & $0.972$ & $62$\\
      $1.0$ & 8.3 & $0.994$ & $88$\\
      $>1.0$ & 0 & \multicolumn{2}{}\\
    \end{tabular}
  \end{center}
\end{table}

Overall, summing up average lengths over the ranges of the experiments, the number of bits generated by our was about
\begin{itemize}
\item $1.015$ times the entropy (1.5~percent redundancy), and
\item $1.0005$ times the corresponding optimal Huffman code (an increase of five hundredths of a percent)
\end{itemize}
This indicates that all but a negligible part of the redundancy is due to the rounding effect of a code tree for a non-dyadic probability distribution, and not to inferior behavior in relation to Huffman coding.

Finally, the total number of bits was $0.737$ times the corresponding Golomb codes, i.e., a 26~percent reduction.
This may appear to be a surprisingly large improvement, but consider that the experiment was done for relatively small $n$ in order to evaluate the redundancy of our code.
For $n\gg m$, the difference between our code and a Golomb code is negligible.

\section{Time Complexity and Efficient Implementation}\label{sec:efficiency}

Computation in encoding and decoding can be separated into three phases. The first computes $m$ and $m''$ when only $p$ is known. This involves floating point operations (logarithm and division), integer ceiling-log, and some integer bitwise and arithmetic operations. The second phase can take place when $n$ is known, and computes $m'$. For the third phase, encoding or decoding of a value $i$ can then take place. Disregarding input/output operations, phases two and three involve division, conditional branching (\emph{if-else}) ceiling-log, bitwise, and arithmetic operations, but no floating point calculations.

\label{sec:timecomplexity}

We assume that arithmetic and bitwise operations (including shift, used for calculating powers of two) are constant-time machine operations. Ceiling-log, $⌈\lg x⌉$, cannot be directly expressed in most programming languages, but if not available as a special operation, it can either be computed by converting to floating point and extracting the exponent bits, or by locating the highest one-bit in the binary representation of~$x-1$ using $\Theta(\lg\lg N)$ operations, where $N$ is an upper bound for $x$. Assuming a fixed word size of at most $64$~bits, $\lg\lg N$ cannot exceed~$6$, and for practical purposes can be regarded as taking constant time. The floating-point logarithm involved in the first phase, however, is less easily dismissed, as it would typically involve a Taylor series computation.

We have generally made the assumption (see section~\ref{sec:introduction}) that computing (and representing) $m$ and $m''$ is rare enough to be considered negligible, while efficiency is critical for phase~2 computation and encoding/decoding.

Consequently, depending on machine model assumptions, time complexity of the critical parts of computation are bounded by either $O(1)$ (arguably true in most practical situations) or $O(\lg\lg N)$ where $N$ is an upper bound for $n$. The time for the first phase, executed a small number of times, takes additional time for two floating point logarithm calculations, typically bounded by $O(w)$, where $w$ is the number of bits of accuracy for the floating-point calculations.

\subsubsection{Branch-Free Implementation}

Classical time complexity analysis assumes that CPU instructions are executed more or less in sequence, where each instruction takes constant time.
The truth for modern processors, where \emph{pipelining} and \emph{out-of-order execution} are crucial to efficiency, is more complex~\cite{ShenLipasti}.
Conditional branches, which the processor may or may not be able to predict, potentially forces execution out of \emph{streaming mode}, and performance suffers.
Consequently, when writing program code without detailed knowledge of code generation and target machine architecture, it is generally a good idea to write programs with as few conditional branching constructs (e.g., \emph{if} and \emph{while} statements) as possible.

With a straightforward implementation of our algorithms, a single encode or decode runs through at least three \emph{if-else} statements. Although constant-time in the sense of classical algorithm analysis, they can be detrimental to practical performance.
Our algorithms do however lend themselves to be implemented \emph{without} any conditional statements, by using some bitwise tricks that take advantage of the two's complement integer representation of modern computers.
The following tricks can partly be regarded as folklore in programming communities. Knuth~\cite{Knuth4A} has examined some related techniques more closely. Our examples use the C programming language.


\begin{figure}[b]
  \vbox{
    \hbox{\includegraphics[scale=.66]{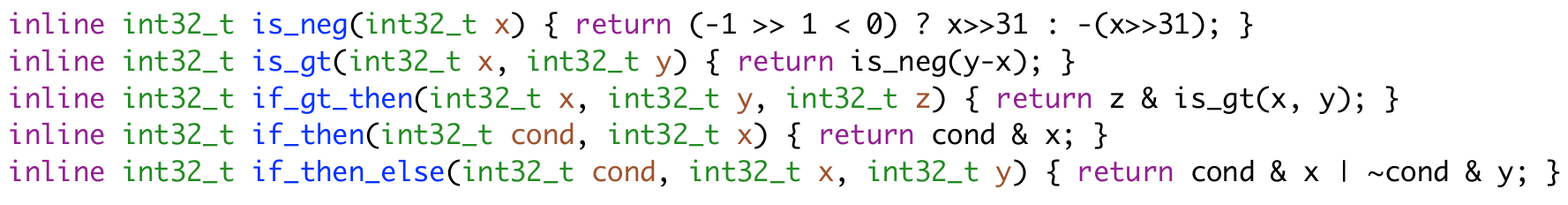}}
    \vspace{-2ex}
    \hbox{\rule{1\textwidth}{.25pt}}
    \vspace{1ex}
    \hbox{\hspace{-.3em}
      \raisebox{1.4\height}{\includegraphics[scale=.66]{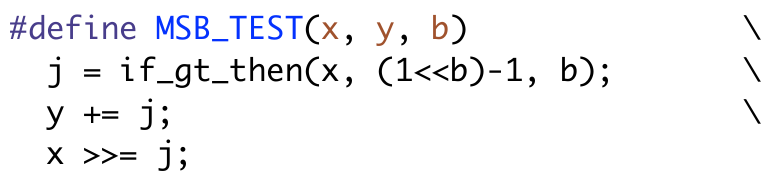}}
      \hspace{.5em}
      \raisebox{.5ex}{\rule{.25pt}{21ex}}
      \hspace{.5em}
      \includegraphics[scale=.66]{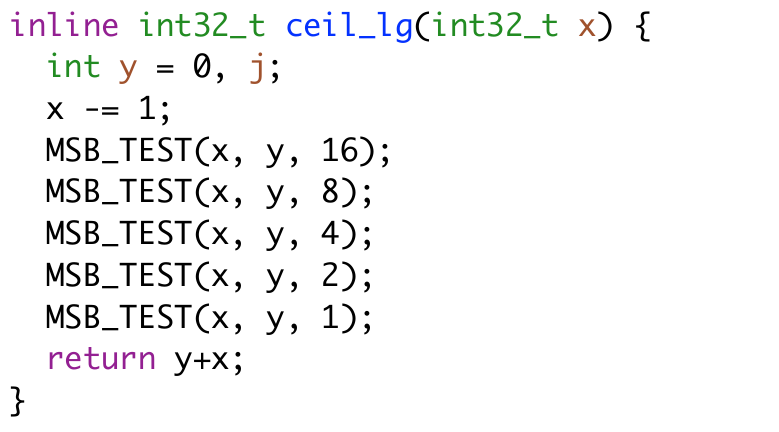}
    }
  }

  \caption{Components for branch-free implementation.} \label{fig:branchfree-tricks}
\end{figure}

In two's complement representation, the leftmost bit is one for any negative number, and zero for any non-negative number. At the core of the tricks is smearing the sign bit across the whole word, producing $-1$ (all one-bits, which we take as our representation of \emph{true}) if the number is negative, and zero (representing \emph{false}) otherwise. Using \emph{arithmetic shift right}, this is done simply by shifting the sign bit $w-1$ positions to the right, where $w$ is the number of bits in a word. For simplicity, we henceforth assume that $w=32$. (Transformation to other word sizes is straightforward.) Hence with, \verb'>>' denoting right shift, the operation is “\verb'x>>31'”. On the other hand, if shifting is \emph{logical}, bits shifted in from the left are zero rather than copies of the sign bit, producing $1$ instead of $-1$, we must negate the result, i.e., “\verb'-(x>>31)'”. As it is not always possible to know whether shifting is arithmetic or logical (the C standard leaves it undefined) we use a conditional expression (see \verb'is_neg' in figure~\ref{fig:branchfree-tricks}), trusting that the compiler will not translate it to a conditional branch, since the value of the condition is known in compile time.


For positive numbers $x$ and $y$, $x>y \Leftrightarrow y-x<0$. Hence, we immediately have a method to compare the numbers that produces $-1$ if $x>y$ and zero otherwise (\verb'is_gt' in figure~\ref{fig:branchfree-tricks}). 
Since $-1$ is represented as all one-bits, a bitwise \emph{and} “\verb'x&c'” preserves \verb'x' if \verb'c'$=-1$ (“true”) and is zero if \verb'c' is zero (“false”). This is useful in itself (see \verb'is_gt_then' in figure~\ref{fig:branchfree-tricks}), and can be expanded to a full conditional expression by the additional use of bitwise \emph{or} and negation (\verb'if_then' and \verb'if_then_else' in figure~\ref{fig:branchfree-tricks}).
As noted in \journal{section~\ref{sec:timecomplexity}}{in the first part of this section},
$⌈\lg x⌉$
can be computed in $O(\lg w)$ tests for whether the most significant bit of $x-1$ is in the left or right half of a range of bits (reminiscent of binary search). Since $\lg 32 = 5$, we can implement integer ceiling-log simply using five tests. (\verb'ceil_lg' in figure~\ref{fig:branchfree-tricks} implements this with the help of the macro \verb'MSB_TEST'.)

\begin{figure}[t]
  \includegraphics[scale=.66]{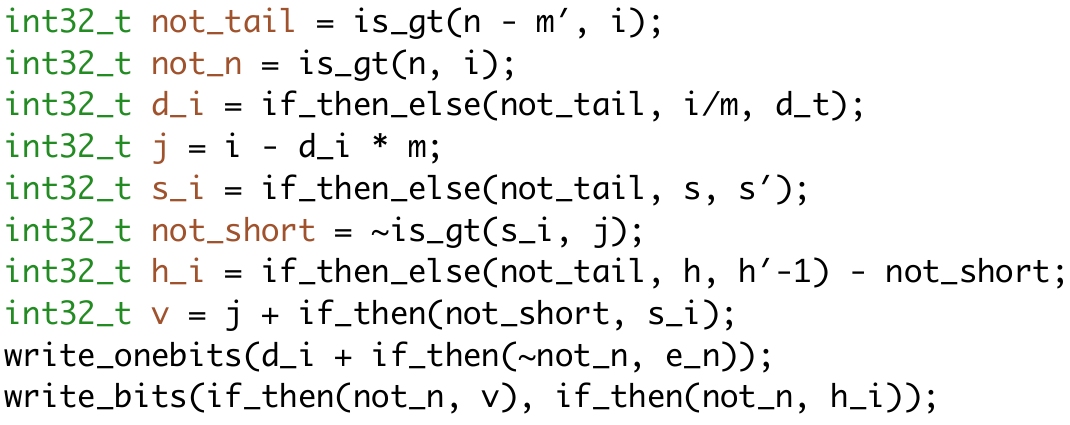}

  \caption{Branch-free implementation of encoding. Values
    \texttt{m}, $\texttt{m}'$,
    \texttt{d\_t}, \texttt{s}, $\texttt{s}'$, \texttt{h}, $\texttt{h}'$, and \texttt{e\_n}
    are assumed computed in the previous phases (not shown).} \label{fig:branchfree-enc}
\end{figure}

Figure~\ref{fig:branchfree-enc} shows an example implementation of encoding (without the two phases of parameter computation). The full branch-free implementation can be found in appendix~2, as well as via \url{http://fgcode.avadeaux.net/}.

It should be noted that the branch-free version is not guaranteed to be the most efficient for all combinations of compilers and target machines. An obvious potential inefficiency is due to branch-free techniques often computing values that are then discarded, since it effectively follows all execution paths of the algorithm.

\section{Conclusion}\label{sec:conclusion}

Adding an efficient coding for the geometric distribution bounded by a potentially small constant to the repertoire of entropy codings, can plausibly contribute to the design of efficient text compression algorithms and compact data structures.


Given that our code is quite simple and addresses a natural situation that arises in practice (from experience by the author of this work), it could be argued that the problem might as well have been solved at least as early as in the 1970s. However, as no code with these properties is previously established in the data compression community, this research fills in a quite literal gap.

\clearpage

\section*{Appendix 1: Straightforward Python Implementation}

\includegraphics[scale=.61]{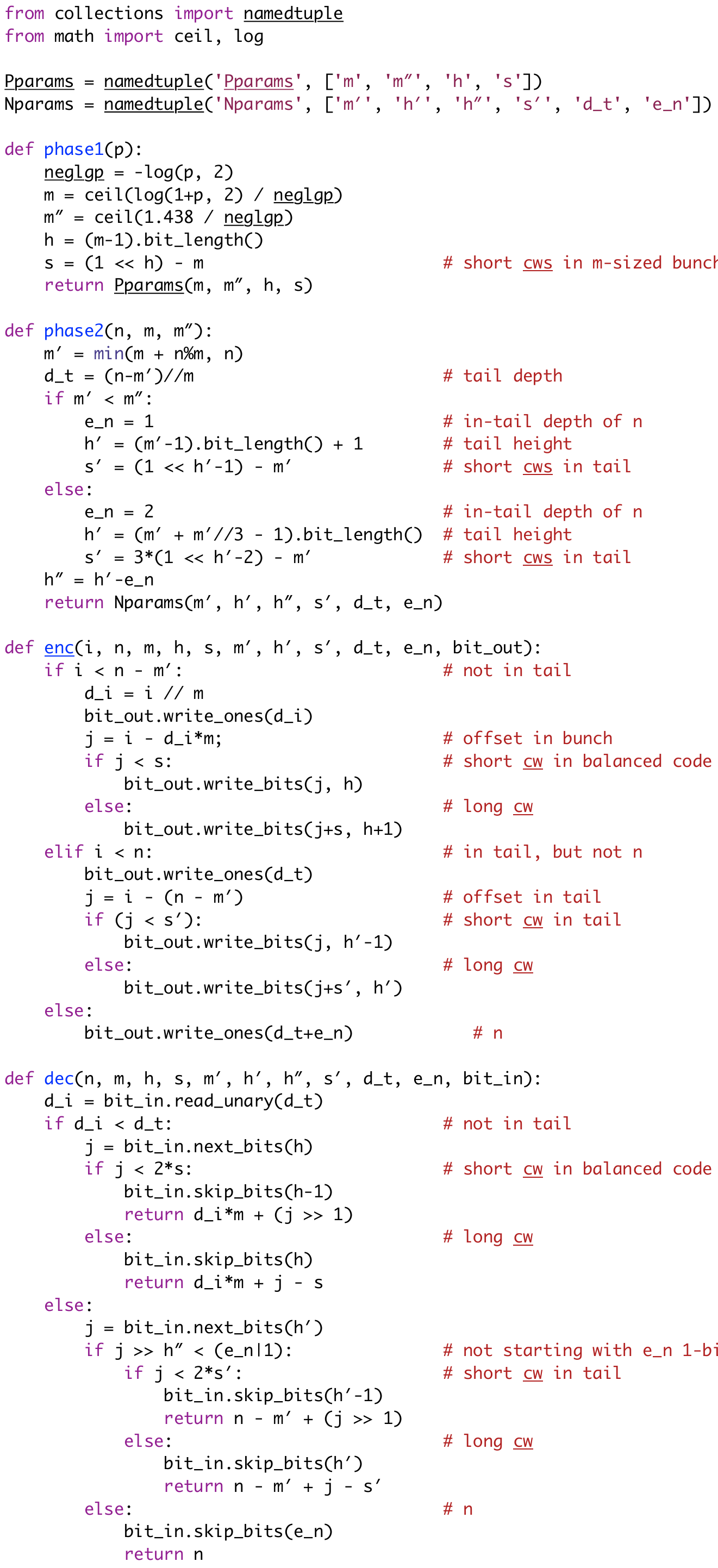}

\section*{Appendix 2: Branch-Free C Implementation}

\includegraphics[scale=.61]{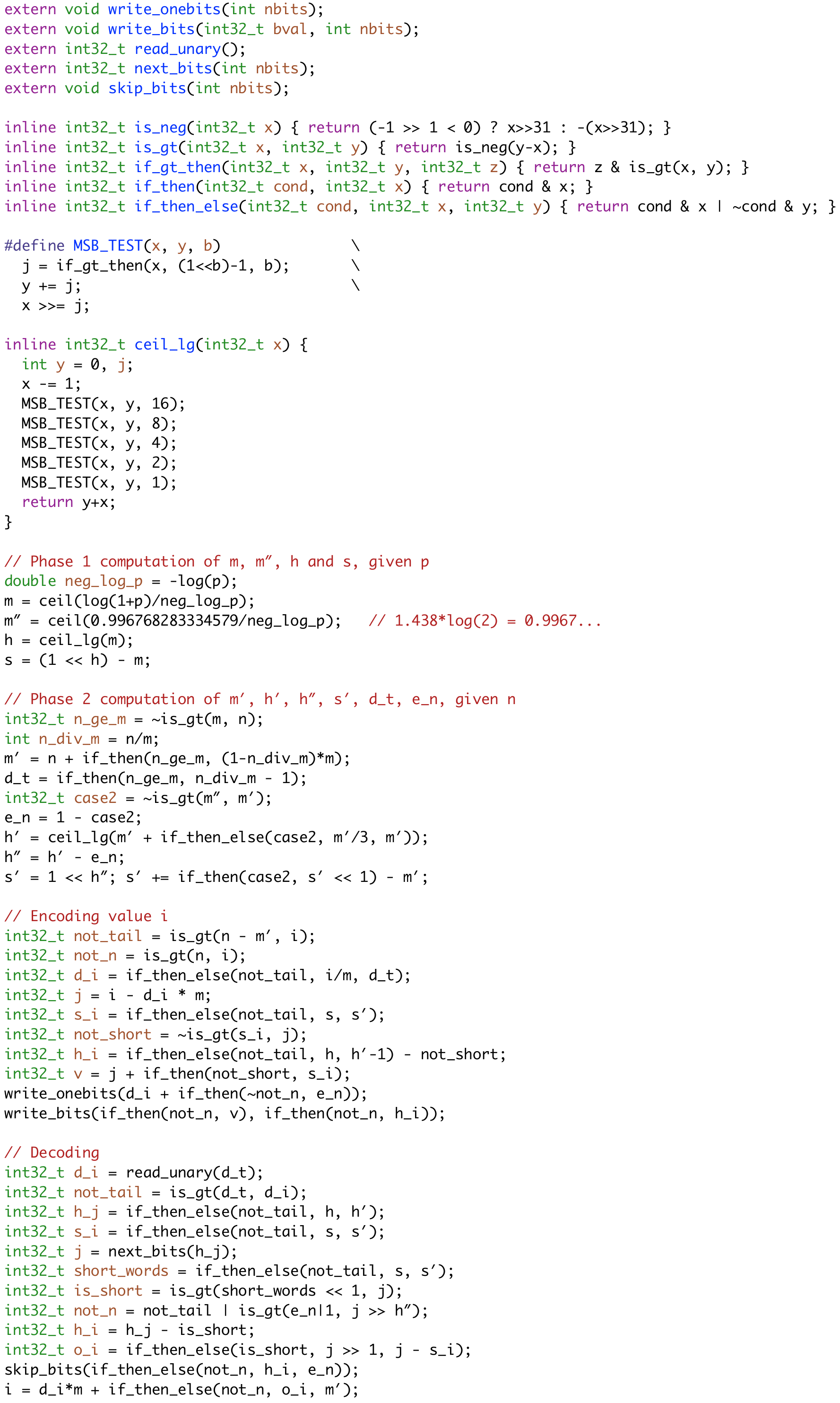}

\bibliographystyle{splncs04}
\bibliography{jesper-general}

\end{document}